\documentclass[a4paper,11pt]{article}
\pdfoutput=1 

\usepackage{jheppub} 

\usepackage[T1]{fontenc} 

\usepackage{slashed,cases,comment,bm,here,colortbl,amsmath,xcolor}

\usepackage{physics}
\usepackage{mathtools}
\usepackage{verbatim}
\usepackage{float}

\usepackage{graphicx}
\usepackage{multirow}
\usepackage{tabularx}
\usepackage{makecell}
\usepackage{booktabs}
\usepackage{wrapfig}

\usepackage{MnSymbol}

\usepackage{framed}
\definecolor{shadecolor}{rgb}{0.95,0.95,0.95}

\title{\boldmath A study of potential non-relativistic QED pairs in the Keldysh-Schwinger formalism}

\author[a]{Tobias Binder, }

\affiliation[a]{Kavli IPMU (WPI),\\ UTIAS, The University of Tokyo, Kashiwa, Chiba 277-8583, Japan}

\emailAdd{tobias.binder@ipmu.jp}

\abstract{Potential non-relativistic Quantum Electrodynamics and the Keldysh-Schwinger formalism is used to derive  Kadanoff-Baym-like equations for two-body field correlators. These cover the out-off-equilibrium dynamics and spectrum of heavy particle-antiparticle pairs under ultra-soft transitions inside a plasma background, whose temperature is much smaller compared to the typical relative pair momentum. It is shown that the dynamical equation for the pair phase-space distribution function, containing recombination and dissociation via the electric dipole interactions, is consistent with the previous open quantum system treatment of bound states in the Coulomb limit of the two-body spectral function.}

\begin{document} 
\maketitle
\flushbottom

\section{Introduction}
\label{sec:intro}

We consider non-relativistic pairs inside a plasma environment with a temperature $T$. Two concrete Physics scenarios are heavy quark-antiquark pairs (Quarkonia) inside the quark-gluon plasma (QGP) and heavy Dark Matter (DM)~(e.g., \cite{Hisano:2006nn,Mitridate:2017izz}) in the early Universe. While in the former case, Quarkonia are often initially produced out-off equilibrium, in the latter case it is the Hubble expansion rate that leads at some point to departure from thermal equilibrium for freeze-out situations. Besides these and differences in the nature of the (hypothetical) particles, the overall theoretical description of the dynamical evolution of the pairs inside a plasma environment shares many similarities.

In particular, in the literature of Quarkonia inside the QGP~\cite{Akamatsu:2014qsa,Brambilla:2016wgg,Brambilla:2017zei,Yao:2018nmy,Yao:2018sgn,Brambilla:2019tpt,Yao:2020xzw,Yao:2020eqy,Akamatsu:2020ypb} and heavy DM in the early Universe~\cite{Binder:2020efn,Binder:2021otw}, the open quantum system treatment has been established. In this treatment, the density-matrix evolution of the pairs are derived starting from the von Neumann equation, which results in a Lindblad equation under some assumptions of the  Hamiltonian and the density matrix of the system. The operators in the Lindblad equation are determined by potential non-relativistic effective theory~\cite{Pineda:1997bj,Brambilla:1999xf,Brambilla:2004jw} (pNREFT) for $T$ much smaller compared to the typical relative momentum of the pair (Bohr momentum).

In this work, we also adopt pNREFT but derive the evolution equations in the Keldysh-Schwinger formalism.
The results are Kadanoff-Baym-like equations for the two-body field correlation functions.
As a difference to the traditional Kadanoff-Baym equations, these correlation functions have two time and \emph{four spatial} arguments since two-body fields are composite operators. This makes the derivation more extended, e.g., the identification of a subset of spatial arguments for which a Wigner transformation and gradient expansion leads to desired results.  

At first place, our interest is to clarify under which approximations one of these coupled differential equations for two-body field correlation functions coincides with a Boltzmann equation for perfectly Coulombic bound states. The latter can be directly derived from the open quantum system treatment by projecting the pNREFT Lindblad equation into the bound state sub-space, see, e.g., Ref.~\cite{Yao:2018nmy}. Such a recovery would give us confidence that Kadanoff-Baym-like equations for correlation functions involving composite operators can indeed reproduce known results.

In general, Kadanoff-Baym equations can deal with quasi particle excitation.
Thus we may expect that our derived equations will also allow for more general excitation of the bound states inside the plasma beyond their Coulomb limit. Our main concern is to identify the structure of the collision term for bound state formation and dissociation including such quasi-particle effects.
 
This work is organized as follows. Section~\ref{sec:pnr} introduces the utilized pNREFT, in particular pNRQED. Based on this EFT and the Keldysh-Schwinger formalism, we derive the evolution equations of two-body field correlation functions in Section~\ref{sec:eom}, including the truncation and closure of the infinite correlator hierarchy. In Section~\ref{sec:open}, assumptions and approximations are clarified, which lead to the previous open quantum system treatment of the bound states. The work is concluded in Section~\ref{sec:con}. 

\section{Potential non-relativistic effective field theory}
\label{sec:pnr}
The relativistic QED Lagrangian under consideration is given by
\begin{align}
\mathcal{L} = i \bar{\chi}\gamma^{\mu}\partial_{\mu}\chi -m_{\chi} \bar{\chi} \chi -g \bar{\chi} \gamma^{\mu} \chi A_{\mu} -\frac{1}{4} F^{ \mu \nu }F_{ \mu \nu } +\mathcal{L}^{\text{env}}[A]. \label{eq:darkQED}
\end{align}
We are interested in the non-equilibrium dynamics of the non-relativistic $\chi$ fields inside a plasma environment with temperature $T$. Under the assumption $v_{\text{rel}}\ll1$, $\alpha=g^2/(4\pi)\ll 1$ and $T$ much smaller than the typical relative momentum ($\sim \alpha m_{\chi}$), the relevant part of the pNRQED Lagrangian~\cite{Pineda:1997bj,Brambilla:1999xf,Brambilla:2004jw} can be written as
\begin{align}
\mathcal{L}^{\text{pNRQED}} &=   \int \text{d}^3 r \; \text{Tr}\{ O^{\dagger} (\mathbf{x},\mathbf{r},t) \left[ i\partial_t - h +  \; \mathbf{r} \cdot g \mathbf{E}(\mathbf{x},t) \right] O(\mathbf{x},\mathbf{r},t)\} -\frac{1}{4} F^{\mu\nu}F_{\mu\nu}+\mathcal{L}^{\text{env}}[A] ,\label{eq:pNRL}
\end{align}
with the trace over spin indices, $O$ being the \emph{two-body field operator} of a ($\chi$) particle-antiparticle pair with reduced mass $m$ and total mass $M$, depending on the center-of-mass coordinate $\mathbf{x}$ and relative distance $\mathbf{r}$, and
\begin{align}
h = -\frac{\nabla_\mathbf{x}^2}{2M} - \frac{ \nabla_\mathbf{r}^2}{2m} + V(r) +\cdots.
\end{align}
The attractive Coulomb potential is given by $V(r)= -\frac{\alpha}{r}$. Higher order terms in the non-relativistic expansion are neglected in the $[\cdots]$ terms and we shall focus in this work on the ultra-soft contributions. In particular, the \emph{electric dipole operator}, $ \mathbf{r}  \cdot g \mathbf{E}$ in the above Lagrangian leads to transitions among the two-body fields, such as bound-state formation and dissociation. The precise content of the plasma environment is left open and will be implicitly contained in the later introduced electric field correlators.

\section{Two-body field correlator on the Keldysh-Schwinger contour}
\label{sec:eom}
To derive the evolution equations, let us start by defining a \emph{two-body field two-point correlation function}, with two time and four spatial arguments and four spin indices, as:
\begin{align}
G_{tss^{\prime}t^\prime}(x,y;\mathbf{r},\mathbf{r}^\prime)\equiv \langle T_\mathcal{C} O_{ts}(x,\mathbf{r}) O^{\dagger}_{s^{\prime}t^\prime}(y,\mathbf{r}^\prime)   \rangle,
\end{align}
where $\langle T_\mathcal{C} ... \rangle \equiv \text{Tr}[ \rho T_\mathcal{C} (...)]$, $\rho$ is the density matrix, and $\mathcal{C}$ denotes time ordering on the Keldysh-Schwinger contour~\cite{Schwinger:1960qe,Keldysh:1964ud} (for an introduction, see, e.g., Ref.~\cite{Binder:2018znk}). From the path integral, we derive the equation of motions (EoMs) for this two-point function with the pNRQED Lagrangian in Eq.~(\ref{eq:pNRL}), by using the measure invariance principle under infinitesimal field shifts, leading to:
\begin{align}
(i\partial_{x^0}-h_x)G_{tss^{\prime}t^\prime}(x,y;\mathbf{r},\mathbf{r}^\prime)&= i\delta_{s s^{\prime}}\delta_{t t^\prime}\delta^4_{\mathcal{C}}(x,y)\delta^3(\mathbf{r}-\mathbf{r}^{\prime}) \label{eq:i1} \\ & - g r_i \langle T_\mathcal{C} E_i(x) O_{ts}(x,\mathbf{r}) O^{\dagger}_{s^{\prime}t^\prime}(y,\mathbf{r}^\prime) \rangle , \nonumber  \\
(-i\partial_{y^0}-h_y^\prime)G_{tss^{\prime}t^{\prime}}(x,y;\mathbf{r},\mathbf{r}^\prime)&= i\delta_{s s^{\prime}}\delta_{t t^\prime}\delta^4_{\mathcal{C}}(x,y)\delta^3(\mathbf{r}-\mathbf{r}^{\prime}) \label{eq:i2} \\ & - g r_i^{\prime} \langle T_\mathcal{C} E_i(y)  O_{ts}(x,\mathbf{r}) O^{\dagger}_{s^{\prime}t^\prime}(y,\mathbf{r}^\prime) \rangle. \nonumber
\end{align}
In this notation, primed $h$ in the second line indicates that the relative position is $r^{\prime}$.
We shall make the spin indices implicit from now on, while keeping in mind that electric dipole interactions are spin conserving. The three point functions with the electric field, as occurring on the right hand side, obey the EoMs:
\begin{align}
(i\partial_{z^0}-\bar{h}_z)\langle T_\mathcal{C} E_i(x) O(z,\bar{\mathbf{r}}) O^{\dagger}(y,\mathbf{r}^{\prime}) \rangle &= i\langle  E_i(x) \rangle \delta^4_{\mathcal{C}}(z,y)\delta^3(\bar{\mathbf{r}}-\mathbf{r}^\prime) \\ & -g \bar{r}_j \langle T_\mathcal{C} E_i(x) E_j(z) O(z,\bar{\mathbf{r}}) O^{\dagger}(y,\mathbf{r}^\prime) \rangle, \nonumber \\
(-i\partial_{z^0}-\bar{h}_z)\langle T_\mathcal{C} E_i(y) O(x,\mathbf{r}) O^{\dagger}(z,\bar{\mathbf{r}}) \rangle &= i\langle E_i(y) \rangle \delta^4_{\mathcal{C}}(x,z)\delta^3(\mathbf{r}-\bar{\mathbf{r}}) \\ & -g \bar{r}_j \langle T_\mathcal{C} E_i(y) E_j(z) O(x,\mathbf{r}) O^{\dagger}(z,\bar{\mathbf{r}}) \rangle. \nonumber
\end{align}
These can be rewritten into integral form as
\begin{align}
\langle T_\mathcal{C} E_i(x) O(z,\tilde{\mathbf{r}}) O^{\dagger}(y,\mathbf{r}^\prime) \rangle &= \langle  E_i(x) \rangle G_0(z,y;\tilde{\mathbf{r}},\mathbf{r}^\prime) \\ & +i g \int_{\mathcal{C}} \text{d}^4w\text{d}^3\bar{r} \; G_0(z,w;\tilde{\mathbf{r}},\bar{\mathbf{r}})  \bar{r}_j \langle T_\mathcal{C} E_i(x) E_j(w) O(w,\bar{\mathbf{r}}) O^{\dagger}(y,\mathbf{r}^\prime) \rangle, \nonumber  \\
\langle T_\mathcal{C} E_i(y)  O(x,\mathbf{r}) O^{\dagger}(z,\tilde{\mathbf{r}}) \rangle &=  \langle  E_i(y) \rangle G_0(x,z;\mathbf{r}, \tilde{\mathbf{r}}) \\ & + i g \int_{\mathcal{C}} \text{d}^4w \text{d}^3\bar{r}  \; \bar{r}_j \langle T_\mathcal{C} E_i(y) E_j(w) O(x,\mathbf{r}) O^{\dagger}(w,\bar{\mathbf{r}}) \rangle G_0(w,z;\bar{\mathbf{r}},\tilde{\mathbf{r}}). \nonumber
\end{align}
Plugging the integral form for the three-point functions into Eq.~(\ref{eq:i1}) and Eq.~(\ref{eq:i2}), leads to:
\begin{align}
(i\partial_{x^0}-h_x)G(x,y;\mathbf{r},\mathbf{r}^\prime)&= i\delta^4_{\mathcal{C}}(x,y)\delta^3(\mathbf{r}-\mathbf{r}^{\prime})-gr_i\langle  E_i(x) \rangle G_0(x,y;\mathbf{r},\mathbf{r}^\prime) \label{eq:main1} \\& -i g^2 \int_{\mathcal{C}} \text{d}^4w \text{d}^3\bar{r} G_0(x,w;\mathbf{r},\bar{\mathbf{r}}) r_i \bar{r}_j \langle T_\mathcal{C} E_i(x) E_j(w) O(w,\bar{\mathbf{r}})O^\dagger(y,\mathbf{r}^\prime)\rangle,  \nonumber \\
(-i\partial_{y^0}-h_y^\prime)G(x,y;\mathbf{r},\mathbf{r}^\prime)&= i\delta^4_{\mathcal{C}}(x,y)\delta^3(\mathbf{r}-\mathbf{r}^{\prime})-gr_i^\prime\langle  E_i(y) \rangle G_0(x,y;\mathbf{r},\mathbf{r}^\prime) \label{eq:main2}\\& -i g^2 \int_{\mathcal{C}} \text{d}^4w  \text{d}^3\bar{r}  \langle T_\mathcal{C} O(x,\mathbf{r}) O^\dagger(w,\bar{\mathbf{r}})  E_j(w) E_i(y) \rangle \bar{r}_j r_i^\prime  G_0(w,y;\bar{\mathbf{r}},\mathbf{r}^\prime).\nonumber
\end{align}

\begin{figure}
    \centering
    \includegraphics[scale=0.6]{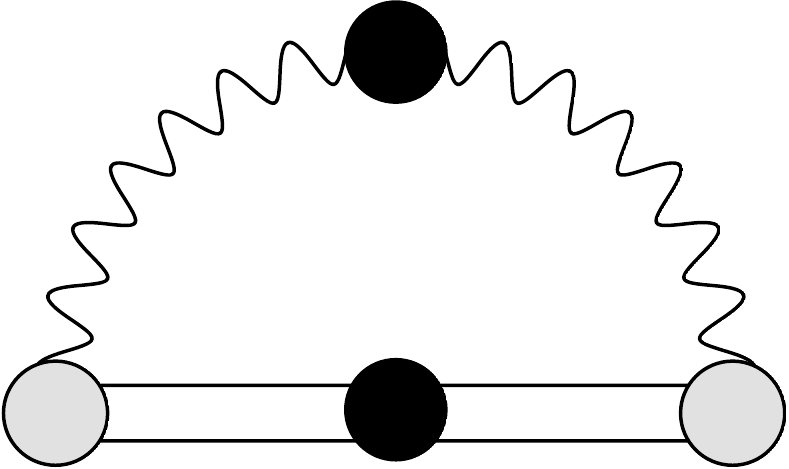}
    \caption{Figure shows a self-energy diagram for the two-body field as a consequence of the leading order approximation of the four point correlators in Eq.~(\ref{eq:main1}) and Eq.~(\ref{eq:main2}). Black blobs represent the interacting electric field and two-body field correlator, respectively. Gray blobs indicate the electric dipole vertices. }
    \label{fig:my_label}
\end{figure}

Eq.~(\ref{eq:main1}) and Eq.~(\ref{eq:main2}) are not closed yet, as expected from the infinitely coupled Martin-Schwinger correlation function hierarchy. In particular, the right hand side of these equations contains an interacting four-point function, which is to leading order the product of the electric field correlator times $G$ correlator. Applying this truncation of the Martin-Schwinger correlator hierarchy to Eq.~(\ref{eq:main1}) and Eq.~(\ref{eq:main2}), gives:
\begin{align}
(i\partial_{x^0}-h_x)G(x,y;\mathbf{r},\mathbf{r}^\prime)&= i \delta^4_{\mathcal{C}}(x,y)\delta^3(\mathbf{r}-\mathbf{r}^{\prime}) -gr_i\langle  E_i(x) \rangle G(x,y;\mathbf{r},\mathbf{r}^\prime) \label{eq:first}  \\&-i  \int_{\mathcal{C}} \text{d}^4w \text{d}^3\bar{r} \; \Sigma_{ij}(x,w;\mathbf{r},\bar{\mathbf{r}}) r_i \bar{r}_j  G(w,y;\bar{\mathbf{r}},\mathbf{r}^\prime), \nonumber \\
(-i\partial_{y^0}-h_y^\prime)G(x,y;\mathbf{r},\mathbf{r}^\prime)&= i \delta^4_{\mathcal{C}}(x,y)\delta^3(\mathbf{r}-\mathbf{r}^{\prime}) -gr_i^\prime\langle  E_i(y) \rangle G(x,y;\mathbf{r},\mathbf{r}^\prime) \label{eq:second} \\& -i  \int_{\mathcal{C}} \text{d}^4w  \text{d}^3\bar{r} \; G(x,w;\mathbf{r},\bar{\mathbf{r}}) \bar{r}_j r_i^\prime   \Sigma_{ji}(w,y;\bar{\mathbf{r}},\mathbf{r}^\prime), \nonumber
\end{align}
where we introduced the two-body field self-energy, $\Sigma_{ij}(x,y;\mathbf{r},\bar{\mathbf{r}}) \equiv g^2 G(x,y;\mathbf{r},\bar{\mathbf{r}})  E_{ij}(x,y)$, with the electric field correlator $E_{ij}(x,y)\equiv \langle T_\mathcal{C} E_i(x)E_j(y)\rangle$. The self-energy is graphically illustrated in Fig.~\ref{fig:my_label}.

To arrive at Eq.~(\ref{eq:first}) and Eq.~(\ref{eq:second}), free two-body field correlators were dressed to account for some higher order terms that are neglected otherwise from the truncation. In relativistic QED, the dressing of the correlators after truncation allows to include the contribution from, e.g., Compton scattering. A more rigorous error estimate in the context of pNRQED requires some mathematical consideration, which we leave for future work. We checked that the approximations made so far conserve the total particle number.

\section{Recovering the open quantum system treatment for bound states}
\label{sec:open}

We turn now to a derivation of a quantum Boltzmann equation for quasi-particle pairs, starting from the coupled Keldysh-Schwinger correlators in Eq.~(\ref{eq:first}) and  Eq.~(\ref{eq:second}).
As a first step, expectation values of the electric field, $\langle E \rangle $, are set to zero. Within this assumption, it is useful to introduce Wigner coordinates for the \emph{Center-of-Mass} (CoM) coordinates for all correlators as:
\begin{align}
T=(x^0+y^0)/2 \;&, t=(x^0-y^0) \;,\\
\mathbf{R}=(\mathbf{x}+\mathbf{y})/2 \;&, \mathbf{l}=(\mathbf{x}-\mathbf{y})\;.
\end{align}
The statistical component $G^{+-}=\langle O^{\dagger} O \rangle $ of Eq.~(\ref{eq:first}) and  Eq.~(\ref{eq:second}) is of particular interest, since it contains the normal ordering of the two-body fields which is related to the occupation number of the pairs. Fourier transforming and gradient expanding the Wigner coordinates, and \emph{subtracting} the $+-$ component of  Eq.~(\ref{eq:second}) from Eq.~(\ref{eq:first}), results in:
\begin{align}
& \left(\partial_T+\frac{\mathbf{p}_{\text{cm}}}{M}\cdot \vec{\nabla}_{\mathbf{R}} \right) \int\text{d}^3r \; G^{+-}(T,\mathbf{R},P;\mathbf{r},\mathbf{r}) =  \label{eq:stat}\\&-\int \text{d}^3r \text{d}^3\bar{r} \; r_i \bar{r}_j\left[\Sigma^{-+}_{ij}(T,\mathbf{R},P;\mathbf{r},\bar{\mathbf{r}})G^{+-}(T,\mathbf{R},P;\bar{\mathbf{r}},\mathbf{r})-\Sigma^{+-}_{ij}(T,\mathbf{R},P;\mathbf{r},\bar{\mathbf{r}})G^{-+}(T,\mathbf{R},P;\bar{\mathbf{r}},\mathbf{r}) \right],\nonumber
\end{align}
where $P=(P^0,\mathbf{p}_{\text{cm}})$ denotes the energy and momentum associated with the Fourier transform of the microscopic variables $t$ and $\mathbf{l}$, respectively. Regarding the relative coordinates, we have taken the limit $\mathbf{r}^{\prime} \rightarrow \mathbf{r}$, such that the potential on the left hand side, as well as the Laplacian for relative coordinates canceled, and finally integrated over the remaining relative position dependence. The resulting kinetic term is of Vlasov type. 

The plasma environment is assumed to be in thermal equilibrium, implying that some of the electric field correlators are related by the Kubo-Martin-Schwinger (KMS) relation as: $E_{ij}^{+-}(P)=f_\gamma^{\text{eq}}(P^0)E_{ij}^{\rho}(P)$,
and $E_{ij}^{-+}(P)=\left[1+f_\gamma^{\text{eq}}(P^0)\right]E_{ij}^{\rho}(P)$, where $f_\gamma^{\text{eq}}$ is the Bose-Einstein distribution and $E^{\rho}_{ij}$ denotes the spectral function of the electric field correlator. 

To describe the dynamics of the quasi-particle pair distribution inside this plasma environment, we take a Kadanoff-Baym-like Ansatz as: $G^{+-}(T,\mathbf{R},P;\mathbf{r},\mathbf{r}^\prime)=  f(T,\mathbf{R},M+P^0;\mathbf{r},\mathbf{r}^\prime) G^{\rho}(T,\mathbf{R},P;\mathbf{r},\mathbf{r}^\prime)$. In contrast to the plasma environment, it is assumed that the pairs are dilute, i.e., $f \ll 1$ such that $G^{-+}(T,\mathbf{R},P;\mathbf{r},\mathbf{r}^\prime)\simeq G^{\rho}(T,\mathbf{R},P;\mathbf{r},\mathbf{r}^\prime)$.

With these relations for the electric and two-body field correlators, the statistical Eq.~(\ref{eq:stat}) can be expressed in terms of the distribution and spectral functions as:
\begin{align}
&\left(\partial_T+\frac{\mathbf{p}_{\text{cm}}}{M}\cdot \vec{\nabla}_{\mathbf{R}} \right)   \int \text{d}^3r \; f(T,\mathbf{R},M+P^0;\mathbf{r},\mathbf{r}) G^{\rho}(T,\mathbf{R},P;\mathbf{r},\mathbf{r}) = \label{eq:main} \\ & - \int   \frac{\text{d}^4K}{(2\pi)^4} \int \text{d}^3r \int \text{d}^3\bar{r} \; G^{\rho}(T,\mathbf{R},K;\mathbf{r},\bar{\mathbf{r}}) g^2 r_i\bar{r}_jG^{\rho}(T,\mathbf{R},P;\bar{\mathbf{r}},\mathbf{r}) E_{ij}^{\rho}(P-K) \times \nonumber \\ & \bigg\{ f(T,\mathbf{R},M+P^0;\bar{\mathbf{r}},\mathbf{r})\left[1+f_\gamma^{\text{eq}}(P^0-K^0)\right]-f(T,\mathbf{R},M+K^0;\mathbf{r},\bar{\mathbf{r}})f_\gamma^{\text{eq}}(P^0-K^0) \bigg\}.\nonumber
\end{align}
The electric field correlator encodes interactions with the plasma environment at next-to-leading order in the coupling expansion. From the full set, Eq.~(\ref{eq:first}) and  Eq.~(\ref{eq:second}), one can derive the retarded and advanced correlator which determines the spectral function of the two-body fields: $G^{\rho}= G^R-G^A$. These are shared in Appendix~\ref{app:retarded} and given by Eq.~(\ref{eq:qdynamics}) and Eq.~(\ref{eq:retarded}) for the macroscopic and microscopic Wigner coordinates, respectively. 

A Boltzmann equation for the bound states has been derived in the open quantum system treatment in, e.g., Ref.~\cite{Yao:2018nmy}. In order to show that Eq.~(\ref{eq:main}) can be made compatible with this bound state Boltzmann equation, we shall finally neglect self-energy corrections to the retarded and advanced two-body field correlators.
In such a case, the resulting two-body spectral function is independent of $T$ and $\mathbf{R}$, and depends on the combination $E \equiv P^0- \mathbf{p}^2_{\text{cm}}/(2M)$. Moreover, the only remaining interaction is the static Coulomb potential. In this approximation, the solution for the retarded and advanced two-body field correlators are given in Eq.~(\ref{eq:retadvcoulomb}), from which we obtain the Coulomb spectral function:
\begin{align}
G^{\rho}_0(E,\mathbf{r},\mathbf{r}^{\prime}) &= (2\pi) \sum_{\mathcal{B}}\delta(E-E_{\mathcal{B}})\psi_{\mathcal{B}}(\mathbf{r})\psi^\star_{\mathcal{B}}(\mathbf{r}^{\prime}) + \theta(E) \sqrt{E} \frac{(2m)^{3/2}}{2\pi} \psi_{E}(\mathbf{r})\psi_{E}^{\star}(\mathbf{r}^{\prime}).\label{eq:Cspec}
\end{align}
The spectrum consists of a sum over discrete and continuous states, where $E_{\mathcal{B}}$ is the (negative) binding energy of a specific bound state with quantum numbers $\mathcal{B}$. The $\psi$ functions are equivalent to the complete and orthonormal set of Coulomb Schr\"odinger wave functions.

Due to the energy gap in the discrete part of the Coulomb spectrum, one can project Eq.~(\ref{eq:main}) into a specific bound state.
In particular, the projection of the left hand side gives:
\begin{align}
&\int_{E_{\mathcal{B}}-\delta}^{E_{\mathcal{B}}+\delta}\frac{\text{d}E}{(2\pi)}\left(\partial_T+\frac{\mathbf{p}_{\text{cm}}}{M}\cdot \vec{\nabla}_{\mathbf{R}} \right)   \int \text{d}^3r \; f(T,\mathbf{R},M+E+\mathbf{p}^2_{\text{cm}}/2M) G^{\rho}_0(E;\mathbf{r},\mathbf{r})= \nonumber \\ & \left(\partial_T+\frac{\mathbf{p}_{\text{cm}}}{M}\cdot \vec{\nabla}_{\mathbf{R}} \right)f_{\mathcal{B}},\label{eq:proj} 
\end{align}
where $f_{\mathcal{B}}\equiv f(T,\mathbf{R},M+E_{\mathcal{B}}+\mathbf{p}^2_{\text{cm}}/2M)$. We used the normalization of the bound state wave functions, $1= \int \text{d}^3r |\psi_{\mathcal{B}}(r)|^2$, and assumed that the phase-space distribution for the bound states does not depend on the relative position. Note that this assumption needs to be relaxed if one considers instead scattering states. The tolerance, $\delta$, is such that the integration range encloses only one specific bound state delta peak in the negative energy spectrum.

Applying the same projection to the right hand side of Eq.~(\ref{eq:main}) and using Eq.~(\ref{eq:Cspec}), we finally obtain a Boltzmann equation for a specific bound state, which reads for bound-scattering state transitions:
\begin{align}
\left(\partial_T+\frac{\mathbf{p}_{\text{cm}}}{M}\cdot \vec{\nabla}_{\mathbf{R}} \right) f_{\mathcal{B}}= \mathcal{C}^{\text{rec}} - \mathcal{C}^{\text{dis}}.\label{eq:boltzmanneq}
\end{align}
The recombination and dissociation term can be expressed, after some shifts and relabeling of momenta, as:
\begin{align}
\mathcal{C}^{\text{rec}}&= \int \frac{\text{d}^3 k_{\text{rel}}}{(2\pi)^3} \int \frac{\text{d}^3 q}{(2\pi)^3} \frac{g^2}{3} |\bra{\psi_{\mathbf{k}_{\text{rel}}}}\mathbf{r}\ket{\psi_{\mathcal{B}}}|^2   E^{\rho}(\Delta E,\mathbf{q}) \\ & \times f(M +\mathbf{k}^2_{\text{rel}}/(2m)+(\mathbf{p}_{\text{cm}}-\mathbf{q})^2/(2M))\left[1+f_\gamma^{\text{eq}}(\Delta E)\right],\nonumber\\
\mathcal{C}^{\text{dis}}&= \int \frac{\text{d}^3 k_{\text{rel}}}{(2\pi)^3} \int \frac{\text{d}^3 q}{(2\pi)^3} \frac{g^2}{3} |\bra{\psi_{\mathbf{k}_{\text{rel}}}}\mathbf{r}\ket{\psi_{\mathcal{B}}}|^2   E^{\rho}(\Delta E,\mathbf{q}) \\ & \times f_{\mathcal{B}}(M +E_{\mathcal{B}}+\mathbf{p}_{\text{cm}}^2/(2M))f_\gamma^{\text{eq}}(\Delta E),\nonumber
\end{align}
where the introduced energy difference, $\Delta E = \mathbf{k}_{\text{rel}}^2/(2 m) - E_{\mathcal{B}} $, is a positive quantity.

$\mathcal{C}^{\text{dis}}$ is consistent with Eq.~(28) in Ref.~\cite{Yao:2018nmy} (after adjusting group theory factors for switching from singlet-octet transitions in QCD to QED pairs and taking in our result the leading order of the electric field correlator). For a relative position independent scattering state distribution function, $\mathcal{C}^{\text{rec}}$ is compatible with Eq.~(29) in the same reference. Integrating Eq.~(\ref{eq:boltzmanneq}) over $\mathbf{p}_{\text{cm}}$, assuming kinetic equilibrium and a homogenous and isotropic distribution (independent of $R$), the resulting number density equation and the thermally averaged bound-state formation cross section with the interacting electric field correlator is consistent with the results obtained in Ref.~\cite{Binder:2020efn,Binder:2021otw}.

\section{Conclusion}
\label{sec:con}

Potential non-relativistic QED was studied in the Keldysh-Schwinger formalism.
In this framework, we derived coupled differential equations for two-body field correlators given in Eq.~(\ref{eq:first}) and Eq.~(\ref{eq:second}), which determine the out-off-equilibrium dynamics of particle-antiparticle pairs under electric dipole interactions. As one of the main results, we have shown that one can indeed recover from those the previous open quantum system treatment of bound states, as given in form of a Boltzmann equation in Eq.~(\ref{eq:boltzmanneq}). Thereby we clarified the underlying assumptions and approximations needed from the viewpoint of our approach.

One assumption which lead to consistency with previous works was that the phase-space distribution of the bound states is independent of the relative distance. In contrast, we observed that a relative position dependent phase-space distribution is required for the two-body scattering states. Such a dependence is expected from the Quarkonium open quantum system treatment of the scattering states, see, e.g., Ref.~\cite{Akamatsu:2020ypb}, and
would be an interesting future study to investigate its implications on the Dark Matter relic abundance.

Another important result is that the collision term for ultra-soft transitions was expressed in terms of interacting correlation functions in Eq.~(\ref{eq:main}). This allows to study higher order corrections to, e.g., bound state formation and dissociation in a thermal field theoretical framework. One of the correlators is the electric field spectral function, which has been already computed inside a plasma environment consisting of ultra-relativistic fermions in, e.g., in Ref.~\cite{Burnier:2010rp,Binder:2020efn,Binder:2021otw}. The novel correlators in the collision term are the two-body spectral functions. In the Coulomb limit (neglecting self-energy corrections), we have shown that these consistently lead to the previous dipole overlap integrals with the usual Schr\"odinger wave functions for scattering and bound states.

Beyond the Coulomb limit, a two-body spectral function description is of advantage in particular if the effective potential is non-Hermitian, leading to a finite energy width of the bound states. This is phenomena is known to occur for top quark pairs even in vacuum~\cite{PhysRevD.43.1500} or if the plasma in-medium effects lead to a thermal width~\cite{Laine:2006ns,Kim:2016kxt,Biondini:2017ufr}. The self-energy contributions in our retarded correlator could also lead to real and imaginary part corrections. From the viewpoint of Refs.~\cite{Brambilla:2008cx,Brambilla:2013dpa}, thermal corrections to the spectrum in the assumed hierarchy of scales ($T \ll \alpha m_{\chi}$) may however be only subleading compared to the ones arising from the electric field correlator in the collision term.

Nevertheless, the assumed hierarchy of scales can only be satisfied for a subset of Coulomb bound states inside a plasma environment. Indeed, the typical relative momentum can be smaller than temperature for some of the excited bound states, implying that a part of the energy spectrum is always melted by mentioned thermal width effects. 
While this shows that we have not addressed a complete thermodynamical description for the full two-particle energy spectrum yet, we would nevertheless put forward the Keldysh-Schwinger description for pairs inside a plasma environment. By properly adapting the initial Lagrangian, one may also account for thermal width effects in the two-particle spectrum. Starting from the pNRQED Lagrangian, we have so far shown that a Keldysh-Schwinger formulation can at least be seen as an alternative derivation of known evolution equations for the regime where $T$ is much smaller compared to the typical relative momentum.

\acknowledgments

T.B. was supported by World Premier International Research Center Initiative (WPI), MEXT, Japan, by the JSPS Core-to-Core Program Grant Number JPJSCCA20200002, by the JSPS KAKENHI Grant Number 20H01895 and JSPS KAKENHI Grant Number JP21H05452.

\appendix

\section{Retarded two-body field correlators}
\label{app:retarded}

Equations for the retarded two-body field correlator, $G^R=G^{++}-G^{+-}$, are derived from Eq.~(\ref{eq:main2}) and Eq.~(\ref{eq:main1}). It is assumed that the expectation value of the electric field vanishes, which allows to solve the equations in terms of Wigner coordinates. 

\emph{Subtracting} the retarded correlator of Eq.~(\ref{eq:main2}) from Eq.~(\ref{eq:main1}), leads to the dynamical retarded equation:
\begin{align}
&\left( i\partial_T +i\frac{\mathbf{p}_{\text{cm}}}{M} \cdot \vec{\nabla}_{\mathbf{R}} + \frac{\nabla_\mathbf{r}^2 - \nabla_{\mathbf{r}^\prime}^2  }{2m} - \left[V(r)-V(r^\prime)\right] \right) G^{R}(T,\mathbf{R},P;\mathbf{r},\mathbf{r}^\prime) = \label{eq:qdynamics}\\ & -i \int \text{d}^3\bar{r} \;   \left[  \Sigma^{R}_{ij}(T,\mathbf{R},P;\mathbf{r},\bar{\mathbf{r}}) r_i \bar{r}_j G^{R}(T,\mathbf{R},P;\bar{\mathbf{r}},\mathbf{r}^\prime) - G^{R}(T,\mathbf{R},P;\mathbf{r},\bar{\mathbf{r}}) \bar{r}_j r_i^\prime \Sigma^{R}_{ij}(T,\mathbf{R},P;\bar{\mathbf{r}},\mathbf{r}^\prime) \right]. \nonumber
\end{align}

\emph{Adding} the retarded correlator of Eq.~(\ref{eq:main2}) and Eq.~(\ref{eq:main1}), leads to the constraint retarded equation:
\begin{align}
&\left(2P^0 + \frac{-2\mathbf{p}^2_{\text{cm}} +  \partial_{\mathbf{R}}^2/2}{2M} + \frac{\nabla_\mathbf{r}^2 + \nabla_{\mathbf{r}^\prime}^2  }{2m} - \left[V(r)+V(r^\prime)\right]\right) G^{R}(T,\mathbf{R},P;\mathbf{r},\mathbf{r}^{\prime})   = i2\delta^3(r-r^{\prime}) \nonumber \\ & -i \int \text{d}^3\bar{r} \;   \left[  \Sigma^{R}_{ij}(T,\mathbf{R},P;\mathbf{r},\bar{\mathbf{r}}) r_i \bar{r}_j G^{R}(T,\mathbf{R},P;\bar{\mathbf{r}},\mathbf{r}^\prime) + G^{R}(T,\mathbf{R},P;\mathbf{r},\bar{\mathbf{r}}) \bar{r}_j r_i^\prime \Sigma^{R}_{ij}(T,\mathbf{R},P;\bar{\mathbf{r}},\mathbf{r}^\prime) \right]. \label{eq:retarded}
\end{align}
Neglecting all self-energy corrections, the only non-trivial equation is Eq.~(\ref{eq:retarded}), which can also be expressed in this approximation as
\begin{align}
\left(E \pm i \epsilon + \frac{\nabla_\mathbf{r}^2}{2m} - V(r)\right) G^{R/A}_0(E;\mathbf{r},\mathbf{r}^{\prime})   &= i\delta^3(r-r^{\prime}),\label{eq:rac1} \\
\left(E \pm i \epsilon + \frac{\nabla_{\mathbf{r}^{\prime}}^2}{2m} - V(r^\prime)\right) G^{R/A}_0(E;\mathbf{r},\mathbf{r}^{\prime})   &= i\delta^3(r-r^{\prime}),\label{eq:rac2}
\end{align}
where we introduced the non-trivial remaining dependence as $E \equiv P^0-\mathbf{p}^2_{\text{cm}}/(2M)$, and made retarded and advanced boundary conditions explicit.
A solution of these Coulomb retarded and advanced correlators is given by:
\begin{align}
G^{R/A}_0(E,\mathbf{r},\mathbf{r}^{\prime})&= i\sum_{\mathcal{B}}\frac{\psi_{\mathcal{B}}(\mathbf{r})\psi_{\mathcal{B}}^{\star}(\mathbf{r}^{\prime})}{E-E_{\mathcal{B}} \pm i \epsilon} +i \frac{(2m)^{3/2}}{2\pi} \int_0^{\infty}  \frac{\text{d}\bar{E}}{(2\pi)} \sqrt{\bar{E}} \frac{\psi_{\bar{E}}(\mathbf{r})\psi_{\bar{E}}^{\star}(\mathbf{r}^{\prime})}{E-\bar{E} \pm i \epsilon},\label{eq:retadvcoulomb}
\end{align}
where the $\psi$ fields solve the homogenous equations (Schr\"odinger equation). Using the standard completeness relation, $\sumint_{\nu} \psi_{\nu}(\mathbf{r})\psi_{\nu}^{\star}(\mathbf{r}^\prime)= \delta^3(r-r^{\prime})$, one can see that Eq.~(\ref{eq:retadvcoulomb}) solves Eq.~(\ref{eq:rac1}) and Eq.~(\ref{eq:rac2}). The spectral correlation function $G^\rho$, entering Eq.~(\ref{eq:main}), is related to the retarded and advanced components as: $G^\rho= G^R-G^A$. For the Coulomb case, which follows from Eq.~(\ref{eq:retadvcoulomb}), the spectral correlator is given by Eq.~(\ref{eq:Cspec}).

\bibliographystyle{JHEP}
\bibliography{BSF}

\end{document}